\documentclass[%
 reprint,
superscriptaddress,
 amsmath,amssymb,
 aps,
 twocolumn,
float]{revtex4-1}

\usepackage{graphicx}
\usepackage{dcolumn}
\usepackage{bm}
\usepackage{xcolor}
\usepackage[caption=false]{subfig}
\usepackage{tabularx}
\begin{document}

\newcommand{\Schro}{Schr\"{o}dinger~}

\newcommand{\ket}[2]{\left|#1,#2\right\rangle} 
\newcommand{\bra}[2]{\left\langle#1,#2\right|} 
\newcommand{\me}[4]{\left\langle#1,#2\right|\tilde{\rho}\left|#3,#4\right\rangle} 

\title{
Circumventing Detector Backaction on a Quantum Cyclotron 
}

\author{X. Fan}
 \email{xingfan@g.harvard.edu}
\affiliation{Department of Physics, Harvard University, Cambridge, Massachusetts 02138, USA}
 \affiliation{Center for Fundamental Physics, Northwestern University, Evanston, Illinois 60208, USA}
\author{G. Gabrielse}
 \email{gerald.gabrielse@northwestern.edu}
 \affiliation{Center for Fundamental Physics, Northwestern University, Evanston, Illinois 60208, USA}
\date{\today}

\begin{abstract}

Detector backaction can be completely evaded when the state of a one-electron quantum cyclotron is detected, but it nonetheless significantly broadens the quantum-jump resonance lineshapes from which the cyclotron frequency can be deduced.  This limits the accuracy with which the electron magnetic moment can be determined to test the standard model's most precise prediction.  A steady-state solution to a master equation, the first quantum calculation for the open quantum cyclotron system, illustrates a method to circumvent the detection backaction upon the measured frequency.          

\end{abstract}

\maketitle

The electron magnetic moment in Bohr magnetons, determined to $3$ parts in $10^{13}$, is the most precisely determined property of an elementary particle \cite{HarvardMagneticMoment2008,HarvardMagneticMoment2011}. A better measurement is currently of great interest because of an intriguing, 2.4 standard deviation discrepancy  \cite{Atoms2019TowardImprovedMeasurement,MullerAlpha2018} with the most precise prediction \cite{atomsTheoryReview2019} of the Standard Model of particle physics (SM).   Tests of the prediction are critical  because these would check important elements of the SM.  These include Dirac theory \cite{DiracTheoryOriginal}, quantum electrodynamics through the tenth order \cite{QED_C8_Lapo, QED_C10_nio, atomsTheoryReview2019}, hadronic contributions \cite{HadronicContribution2013, HadronicContribution2014,WeakHadronic} and possible weak interaction effects \cite{WeakHadronic, WeakCondtribution1,WeakCondtribution2,WeakCondtribution3,WeakCondtribution4}. 
 The intriguing discrepancy has stimulated new theoretical investigations into possible physics beyond the SM \cite{gardner2019light,ALightComplexScalarForTheElectronAndMuonAnomalousMagneticMoments,PhysRevD.98.075011,PhysRevD.98.113002,PhysRevD.99.095034}.   


A quantum cyclotron \cite{QuantumCyclotron} is a single trapped electron that occupies only the ground and first excited states of its cyclotron motion.  Measuring the quantum jump rate between these states as a function of drive frequency produces resonance lineshapes from which  the cyclotron and spin frequencies, and then the electron magnetic moment can be deduced.  Quantum non-demolition (QND) detection makes it possible to completely evade detector backaction in determining the quantum state.  However, the QND coupling does not prevent detector backaction from producing a cyclotron lineshape that is broad and asymmetric enough to prevent more accurate measurements of the cyclotron frequency and the electron moment to investigate the current discrepancy.  The lineshape for the other frequency that must be measured to determine a magnetic moment is much less of an obstacle to measurements of an interesting precision, because its intrinsically different shape is much more symmetric  \cite{Review}.  This ``anomaly frequency'' is the difference of the spin and cyclotron frequencies which can be measured instead of the spin frequency to get the magnetic moment more precisely.

In this Letter, a steady-state solution to a master equation illustrates the possibility of circumventing all detector backaction except that from detector zero point motion, despite the axial detection motion being spread over many quantum states.  The extremely narrow and nearly symmetric cyclotron lineshapes that should result are examples of what is well known to enable significant progress in precision resonant frequency measurements.  Even though resonant frequencies can be extracted from broad and asymmetric lines is principle, in practice this causes a susceptibility to systematic uncertainties. The steady-state solution is the first quantum mechanical solution for a damped quantum cyclotron coupled to a detection oscillator via a QND coupling. The predicted lineshapes for this open quantum system \cite{jacobs_2014} are very different from a previous prediction that assumed a classical detection oscillation  \cite{BrownLineshapePRL,BrownLineshape}.  

The Hamiltonian for the quantum cyclotron \cite{LandauQuantization,Review}
with angular cyclotron frequency, $\omega_c$, 
\begin{equation}
H_c    =\hbar\omega_c
\left(a_c^{\dagger}a_c+\tfrac{1}{2}\right), 
\end{equation}
has the form of a simple harmonic oscillator.  The energy eigenvalues are a ladder of equally spaced Landau levels \cite{LandauQuantization}, $\hbar \omega_c(n_c+1/2)$ with $n_c=0,1,2,\cdots$.  The raising and lowering operators, $a_c^{\dagger}$ and $a_c$, in terms of position and momentum operators differs from that for a simple harmonic oscillator, of course, because circular rather than linear motion is described.  For the same reason, the position representation of energy eigenstates $|n_c\rangle$ are associated Laguerre polynomials rather than the Hermite polynomials for a simple harmonic oscillator.  

For detection, the quantum cyclotron is coupled to a harmonic oscillator with a Hamiltonian,
\begin{equation}
H_z = \hbar \omega_z \left(a_z^{\dagger}a_z+\tfrac{1}{2}\right), 
\label{eq:Hz}
\end{equation}
with energy eigenstates $|n_z\rangle$, eigenvalues $\hbar \omega_z (n_z+1/2)$, and $n_z=0,1, \cdots$.
For an electron in the electrostatic quadrupole potential of a Penning trap, this detection  motion is the axial oscillation of the electron along the magnetic field direction. 
The raising and lowering operators, $a_z^{\dagger}$ and $a_z$, in terms of position and momentum operators are in every quantum mechanics textbook, as are the energy eigenstates in the position representation.  

\begin{figure}[]
    \centering
    \includegraphics[width=\the\columnwidth]{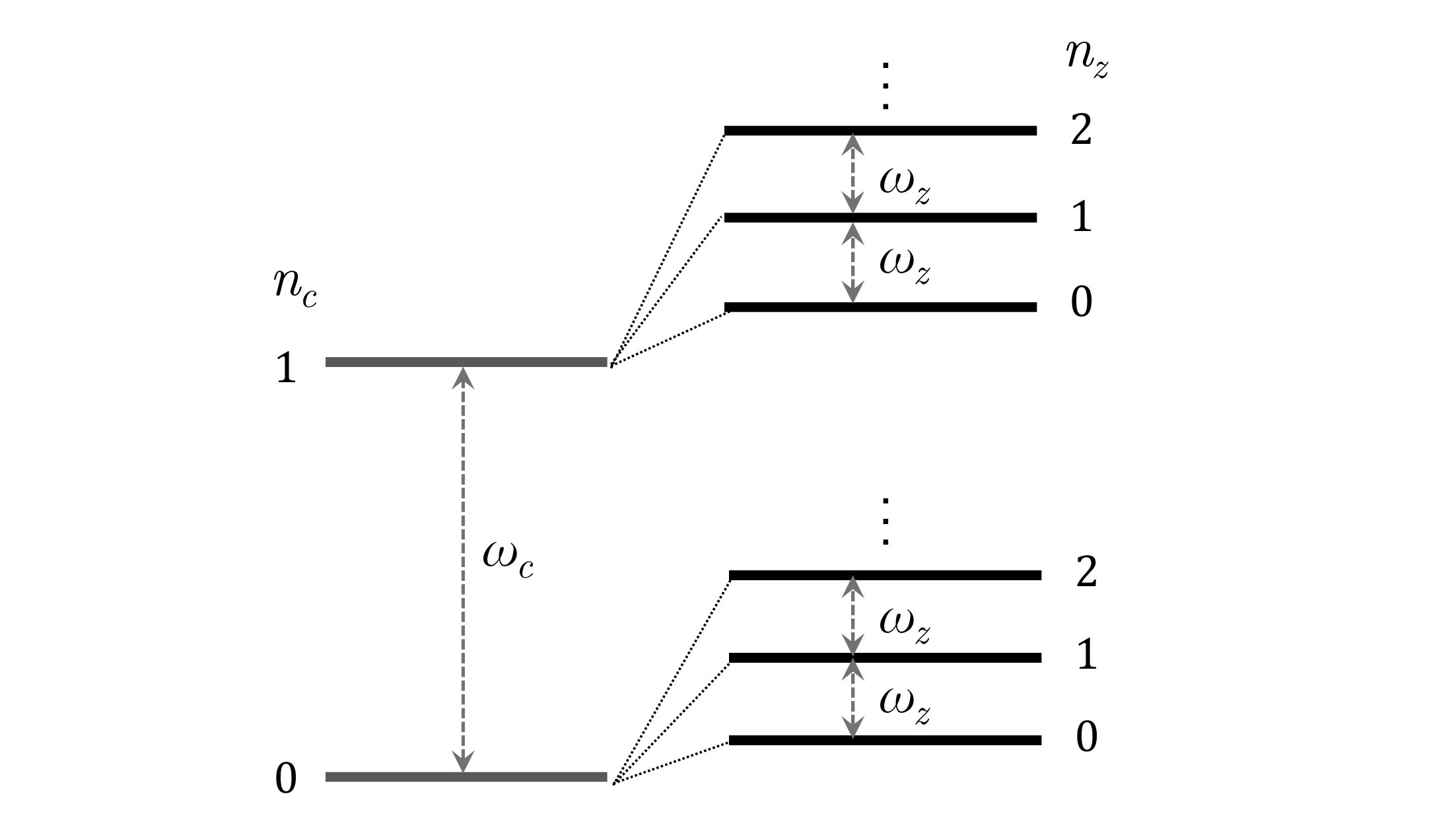}
    \caption{Lowest energy levels for the combined quantum cyclotron and axial detection oscillator (not to scale).}  
    \label{fig:eigenstates}
\end{figure}

The uncoupled Hamiltonian $H_0 = H_c + H_z$ has energy eigenstates $|n_c,n_z\rangle = |n_c\rangle\,|n_z\rangle$, and energy  eigenvalues   
\begin{equation}
E_0(n_c,n_z) = \hbar \omega_c (n_c+\tfrac{1}{2}) + \hbar \omega_z (n_z+\tfrac{1}{2}). \end{equation}
The representation in Fig.~\ref{fig:eigenstates} is not to scale since $\omega_c$ is typically 1000 times larger than $\omega_z$. The magnetron motion present in a laboratory realization of a quantum cyclotron \cite{HarvardMagneticMoment2011} is dropped in our calculation because the frequency scale is smaller by about $\omega_m/\omega_z\approx10^{-3}$ after cooling \cite{Review}. Including magnetron motion would cause negligible broadening and no noteworthy changes.  

Detecting the cyclotron state requires a Hamiltonian
$H=H_c+H_z+V$ with a coupling $V$ of the cyclotron and axial motions.  A small magnetic bottle gradient \cite{DehmeltMagneticBottle} can be added to the uniform field $B\hat{z}$ \cite{Helium3NMR2019} of a Penning trap,
\begin{equation}
    \Delta B =B_2\left( z^2 - \tfrac{1}{2}(x^2+y^2)\right).
\end{equation}
With $\delta_c=\hbar eB_2/(m^2\omega_z)$, the resulting coupling is   
\begin{eqnarray}
V &=&  \hbar\delta_c \left(a_c^{\dagger}a_c+\tfrac{1}{2}\right) (a_z^\dagger a_z+\tfrac{1}{2}),
\label{eq:V} 
\end{eqnarray}
when two rapidly oscillating terms are averaged to zero.   The coupled Hamiltonian has the uncoupled energy eigenstates $|n_c,n_z\rangle$.  The energy eigenvalues  
\begin{eqnarray}
E(n_c,n_z) = E_0(n_c,n_z)
+\hbar \delta_c (n_c+\tfrac{1}{2}) (n_z+\tfrac{1}{2}),    
\end{eqnarray}
acquire a small term that depends upon both $n_c$ and $n_z$.

The coupling $V$ is a QND coupling \cite{QNDScience1980,QNDReview1980, QNDreview1996,1996MarkQND} because it commutes with $H_0$.  The consequence is that detection backaction is completely evaded when the cyclotron quantum state is detected.  This can be seen by writing the energy eigenvalues as 
\begin{equation}
E(n_c,n_z) = \hbar (n_c+\tfrac{1}{2}) + \hbar \widetilde{\omega}_z (n_z+\tfrac{1}{2}). 
\end{equation}
Repeated measurements of the effective axial frequency, 
\begin{equation}
 \widetilde{\omega}_z = \omega_z + \delta_c (n_c+\tfrac{1}{2}), 
 \label{eq:AxialShift}
\end{equation}
will not themselves change the cyclotron state, even as they reveal quantum jumps of the cyclotron state and $n_c$ caused by an external cyclotron driving force.  

Critical to this report is that the QND coupling $V$ that completely evades detection backaction in the determination of the quantum cyclotron state, does not do so for a measurement of $\omega_c$. This can be seen by writing the energy eigenvalues in the alternate form,   
\begin{equation}
E(n_c,n_z) =  \hbar \widetilde{\omega}_c (n_c+\tfrac{1}{2}) + \hbar \omega_z (n_z+\tfrac{1}{2}). 
\end{equation}
Despite the QND coupling, the effective cyclotron frequency, 
\begin{equation}
 \widetilde{\omega}_c = \omega_c + \delta_c (n_z+\tfrac{1}{2}),  
 \label{eq:CyclotronShift}
\end{equation}
shifts in proportion to the axial quantum number. 
This detection backaction shift cannot  be completely evaded because a shift due to axial zero point motion remains even when the axial detection motion is cooled to its $n_z=0$ ground state. Because the shift in this limit is orders of magnitude smaller than what has been attained, the rest of this work focuses upon how this zero-point limit can be attained.  We call this ``circumventing'' detection backaction because the proposal is to achieve this limit while many states beyond $n_z=0$ are populated.  

The $\omega_c$ needed for an electron magnetic measurement must be extracted from 
the resonance lineshape that is the quantum jump rate measured as a function of an external cyclotron drive frequency.  The broad cyclotron linewidth from detection backaction ($\Delta\omega_c/\omega_c=\bar{n}_z\delta_c/\omega_c\approx10^{-9}$ in past experiments \cite{HarvardMagneticMoment2008,HarvardMagneticMoment2011}) limits the accuracy of possible magnetic moment measurements. The distribution of axial states that causes the broad linewidth arises because the axial detection oscillator is weakly coupled to its environment, with a coupling constant, $\gamma_z$.   For times larger than $1/\gamma_z$, this leads to a thermal Boltzmann distribution of axial states.  For a $T=0.1$ K ambient temperature  and $\omega_z/(2\pi)=200$ MHz \cite{HarvardMagneticMoment2008,HarvardMagneticMoment2011}, the average axial quantum number is 
\begin{equation}
\bar{n}_z=\left[\exp\left({\frac{\hbar\omega_z}{k_BT}}\right)-1\right]^{-1} \approx \frac{k_B T}{\hbar \omega_z}\approx 10.
\label{eq:Averagenz}
\end{equation}
For past measurements, the effective axial temperature was actually at least 3 to 5 times higher due to the elevated temperature of the electronics used to detect the axial oscillation and its frequency\cite{HarvardMagneticMoment2011}. 

The cyclotron motion also weakly couples to the thermal reservoir, with a coupling $\gamma_c$. A state $|n_c\rangle$ radiates synchrotron radiation at a rate $n_c \gamma_c$.  In principle, cyclotron states can also absorb blackbody radiation, but at 0.1 K and $\omega_c/(2\pi)=150$ GHz\cite{HarvardMagneticMoment2008}, the number of available blackbody photons is negligible. The average quantum number for a Boltzmann distribution of states is 
 \begin{equation}
     \bar{n}_c=\left[\exp\left({\frac{\hbar\omega_c}{k_BT}}\right)-1\right]^{-1}= 1.2\times10^{-32} \approx 0.
 \end{equation}
 The cyclotron motion  thus  remains in its $n_c=0$ ground state  \cite{QuantumCyclotron} unless a cyclotron driving force is applied.
 
A cyclotron drive adds the Hamiltonian term \begin{equation}
V_c(t) =  \tfrac{1}{2} \hbar \Omega_c \left[a_c^\dagger e^{-i (\omega_c + \epsilon_c) t} + a_c e^{i (\omega_c + \epsilon_c) t} \right].
\label{eq:Drivec}
\end{equation}
The drive strength is given by the angular Rabi frequency, $\Omega_c$, and the drive is detuned from resonance at $\omega_c$ by a detuning $\epsilon_c$. For measurements, the driving force provided by 150 GHz microwaves injected into a trap cavity excites the $|0, n_z\rangle$ states to $|1, n_z\rangle$.
Higher cyclotron states can be neglected because it is less probable to excite from a small population in an excited state, but also because a relativistic shift keeps the cyclotron transitions between excited states off resonance from the drive \cite{Gabrielse85e}.  

A density operator is required for a system that decays and is coupled to a thermal bath. The initial state at time $t=0$ is the cyclotron ground state and a thermal superposition of axial states,
\begin{equation}
\rho(0) = \sum_{n_z=0}^{\infty} 
p_{n_z}(T)~
|0,n_z\rangle \langle 0,n_z|.
\label{eq:InitialState}
\end{equation}
The Boltzmann weighting factors are
\begin{equation}
p_n(T) = \left[1-\exp\left({-\frac{\hbar\omega_z}{k_BT}}\right)\right]\exp\left({-\frac{n\hbar\omega_z}{k_BT}}\right).
\label{eq:pntBoltzman}
    \end{equation}
Explicit calculations show that 150 axial states suffice for axial states in thermal equilibrium at 0.1 K.  

The time evolution of the density operator is described by a Lindblad equation \cite{Lindblad1,Lindblad2,jacobs_2014},
\begin{equation}
\begin{split}
\frac{d\rho}{dt}&=-\frac{i}{\hbar}\left[{H_0} + V + {V}_c,\rho\right] \\
&-\frac{\gamma_c}{2}\left(a_c^\dagger a_c{\rho}-2a_c {\rho} a_c^\dagger+{\rho} a_c^\dagger a_c\right)\\ 
&-\frac{\gamma_z}{2}\bar{n}_z\left(a_za_z^\dagger{\rho}-2a_z^\dagger{\rho} a_z+{\rho} a_za_z^\dagger\right)\\
&-\frac{\gamma_z}{2}\left(\bar{n}_z+1\right)\left(a_z^\dagger a_z{\rho}-2a_z {\rho} a_z^\dagger+{\rho} a_z^\dagger a_z\right).
\end{split}
\label{eq:MasterEquation}
\end{equation}
The first line describes the driven motion. The second describes the incoherent cyclotron decay.  The third and fourth lines describe the incoherent deexcitation and excitation of the axial motion by the thermal bath.  

To efficiently solve the master equation, several transformations are made.  All terms in Eq.~(\ref{eq:MasterEquation}) are transformed to an interaction picture, with  
\begin{equation}
\widetilde{\rho} = e^{i H_0 t /\hbar} \rho  e^{-i H_0 t/\hbar}. 
\end{equation}
Since the coupled system starts and remains axially diagonal, only the probabilities $\widetilde{\rho}_{jk;n_z} = \langle j, n_z | \widetilde{\rho} | k, n_z \rangle$ are needed. The indices $j$ and $k$ are 0 or 1, and $n_z$ takes positive values as large as needed to describe the thermal distribution -- up to about 150 for $\bar{n}_z=10$, as mentioned.  A second transformation, 
\begin{equation}
p_{jk;n_z} \equiv {\rho}_{jk;n_z}e^{i(j-k) \epsilon_c t}
\label{eq:pjk}
\end{equation}
produces a time-independent equation for the $p_{jk;n_z}$, where the detuning $\epsilon_c$ was defined in Eq.~\ref{eq:Drivec}.  The time-dependent probabilities we seek to calculate,
\begin{equation}
 p_{jj;n_z} = \widetilde{\rho}_{jj;n_z} = \langle j,n_z | \rho | j,n_z \rangle  
\end{equation}
are invariant under these transformations.

The master equation in terms of vectors   $\vec{p}_{jk}$, with components  $p_{jk;n_z}$, is
\begin{subequations}
\begin{align}
\frac{d}{dt}\vec{p}_{00}(t)
&=\mathbf{R}(0,0,0)\, \vec{p}_{00}(t)-\Omega_c\textrm{Im}\left[\vec{p}_{01}(t)\right]\nonumber\\&+\gamma_c\vec{p}_{11}(t)\\
\frac{d}{dt}\vec{p}_{01}(t)
&=\mathbf{R}(\epsilon_c,\delta_c,\gamma_c)\, \vec{p}_{01}(t)-i\frac{\Omega_c}{2}(\vec{p}_{11}(t)-\vec{p}_{00}(t))\\
\frac{d}{dt}\vec{p}_{11}(t)
&=\mathbf{R}(0,0,2\gamma_c)\, \vec{p}_{11}(t)+\Omega_c\textrm{Im}\left[\vec{p}_{01}(t)\right].
\end{align}
\label{eq:MatrixEquation}%
\end{subequations}
The nonzero components of the matrices are
\begin{subequations}
\begin{align}
\mathbf{R}(\epsilon_c,\delta_c,\gamma_c)_{n_z,n_z-1} =~ & \gamma_z\bar{n}_z n_z\label{eq:Ra}\\
\mathbf{R}(\epsilon_c,\delta_c,\gamma_c)_{n_z,n_z~~} =~ &   
i\left[-\epsilon_c+(n_z+\tfrac{1}{2})\delta_c  \right]-\tfrac{1}{2}\gamma_c\nonumber\\ 
&-\gamma_z(2\bar{n}_z+1)n_z-\gamma_z\bar{n}_z\label{eq:Rb}\\
 \mathbf{R}(\epsilon,_c\delta_c,\gamma_c)_{n_z,n_z+1} =~ & \gamma_z (\bar{n}_z+1) (n_z+1).
 \label{eq:Rc}
\end{align}
\label{eq:R}%
\end{subequations}
Beside the specified arguments and indices, these equations and matrices depend upon the bath temperature via $\bar{n}_z$, and the axial damping rate, $\gamma_z$.

This vector master equation must be solved for initial conditions (at $t=0$) that $\vec{p}_{00}$ has components $p_{n_z}(T)$ (from Eq.~(\ref{eq:InitialState})) and  $\vec{p}_{01}=\vec{p}_{11}=0$.  The desired resonance lineshape is the probability of a cyclotron excitation,
\begin{equation}
 P = \sum_{n_z=0}^{\infty}  p_{11;n_z} (t_d),  
\end{equation}
is a function of the drive detuning, $\epsilon_c$.  This lineshape depends upon the drive strength, $\Omega_c$ and the time that the drive is applied, $t_d$.   

In general, the vector master equation must be integrated numerically from $t=0$ to $t=t_d$ to determine the lineshape. 
However, for a weak drive with $\Omega_c \ll \gamma_c$ (to avoid power broadening) and  $t_d\gg 1/\gamma_c$ (to let transients damp out), there is a steady state for which the driven cyclotron excitation balances the emission of synchrotron radiation.  This steady-state solution suffices to demonstrate that circumventing detector backaction is possible.

To obtain the steady-state solution, the derivatives in Eq.~(\ref{eq:MatrixEquation}) are set to zero.  The three equations are summed over all axial states and simplified using
\begin{eqnarray}
&&\sum_{n_z=0}^\infty {p}_{00;n_z}(t)\approx  \sum_{n_z=0}^{\infty} p_{n_z}(T) = 1 \\
&&\sum_{n_z=0}^\infty p_{11;n_z}(t) \ll  1\\
&&\sum_{n_z=0}^{\infty}\left( \mathbf{R}(0,0,2\gamma_c)\, \vec{p}_{11} \right)_{n_z}
=-\gamma_c
\sum_{n_z=0}^\infty p_{11;n_z},
\label{eq:p11}
\end{eqnarray}
The first two simplifications pertain for a weak drive and make terms involving $\vec{p}_{11}$  negligible compared to those involving $\vec{p}_{00}$.  The third pertains because $\mathbf{R}(0,0,2\gamma_c)$ has a simple structure and axial damping does not change the total population in states $|1,n_z\rangle$.  The result is the  steady-state probability for cyclotron excitation by a weak drive,  
 \begin{equation}
  P=-\frac{\Omega_c^2}{2\gamma_c}   \textrm{Im}\left[
  \sum_{n_z=0}^{\infty}\left(
  i \mathbf{R}(\epsilon_c,\delta_c,\gamma_c)^{-1}\vec{p}(T)
  \right)_{n_z}\right].  
\label{eq:P}
\end{equation}
The vector $\vec{p} (T)$ has the Boltzmann factors $p_{n_z}(T)$ as its components.  
In the $T=0$ limit, the steady-state lineshape becomes the expected Lorentzian.


Direct numerical integrations of the master equation (Eq.~(\ref{eq:MatrixEquation})) and the steady-state solution in Eq.~(\ref{eq:P}) provide the first fully quantum treatment of the coupled and open cyclotron and axial system.  (More details, including comparisons of direct integrations and steady-state solutions of the master equation, will be published in a longer work that deals with measuring magnetic moments more generally \cite{Fan2020EvadingBackActionPRA}.) The lineshape calculation \cite{BrownLineshape,BrownLineshapePRL} previously available (and used to  predict and analyze all experiments to date) assumed a classical axial oscillation undergoing Brownian motion -- and predicted a very different lineshape. 

We now investigate detector backaction and how it can be circumvented, with estimates first, and then with quantum lineshape calculations.  The result of a thermal distribution of axial states is that a cyclotron drive will make cyclotron transitions over a range of cyclotron drive frequencies, $\Delta \epsilon_c > \bar{n}_z \delta_c$. For the best measurement,  the bath temperature was 0.3 K and above, which corresponds to a spread $\Delta \omega/\omega_c > 800$ ppt (A part per trillion, ppt, is 1 part in $10^{12}$).  Line splitting made it possible to obtain a 300 ppt uncertainty.  Even at 0.1 K temperature, the backaction linewidth will still spread the cyclotron excitation over a broad width. Reducing the coupling strength ($\delta_c$ in Eq.~(\ref{eq:V})) would reduce the backaction.  However, this is not an option because this simultaneously reduces the sensitivity needed to detect the individual states of the quantum cyclotron.

The new possibility proposed here is circumventing backaction by resolving the cyclotron excitations that an electron makes during the time it is in its axial ground state from those made  while the system is in other axial states. Resolving $\delta_c$, the cyclotron frequency shift for axial states with $n_z=0$ and $n_z=1$, requires two conditions,
\begin{eqnarray}
&\delta_c &\gg \gamma_c + 2 \, \bar{n}_z \gamma_z  \label{eq:Condition1}  \\
&\delta_c &\gg 1/t_d \label{eq:Condition2}.
\end{eqnarray}
The first (from the diagonal damping term in Eq.~(\ref{eq:Rb})) requires that the shift be larger than both the cyclotron damping width, $\gamma_c$, and the axial width contribution, $2 \bar{n}_z \gamma_z$.  The latter arises because the underlying physics of the master equation is that probability transfers between the axial oscillation and the thermal reservoir at an average rate going as $\bar{n}_z \gamma_z$. The second requirement is a drive applied long enough that the frequency-time uncertainty principle does not broaden the lineshape.  

The shift $\delta_c/(2\pi) = 4$ Hz used for measurements is much smaller than the extremely small cyclotron damping width, $\gamma_c/(2\pi) = 0.03$ Hz, realized using a microwave cavity to inhibit spontaneous emission \cite{InhibitionLetter}.  At the ambient temperature of experiments, $T = 0.1$ K, this leaves means that $\gamma_z/(2\pi) \ll 0.2$ Hz is needed. This requirement was not met by the $\gamma_z/(2\pi) = 1$ Hz of the best measurement.  Resolving axial quantum structure thus requires reducing $\gamma_z$ by about two orders magnitude.  The second condition (Eq.~(\ref{eq:Condition2})) is met by simply applying the cyclotron drive for much longer than 40 ms.   

The axial damping rate cannot simply be reduced by this large factor because the induced signal needed to deduce the cyclotron state from $\tilde{\omega}_z$ in Eq.~(\ref{eq:AxialShift}) reduces to unusable levels.  One solution would be to rapidly switch between large $\gamma_z$ for cyclotron state detection, and a small  $\gamma_z$ during the time a drive is applied to make cyclotron quantum jumps.  A cryogenic HEMT switching circuit that operates with essentially no power dissipation was recently developed and demonstrated for this purpose \cite{FanRFSwitch2020Arxiv}.  For our estimates and calculations the large and small $\gamma_z$ realized in the lab demonstration are used.

\begin{figure}[htbp!]
    \centering
    \includegraphics[width=\the\columnwidth]{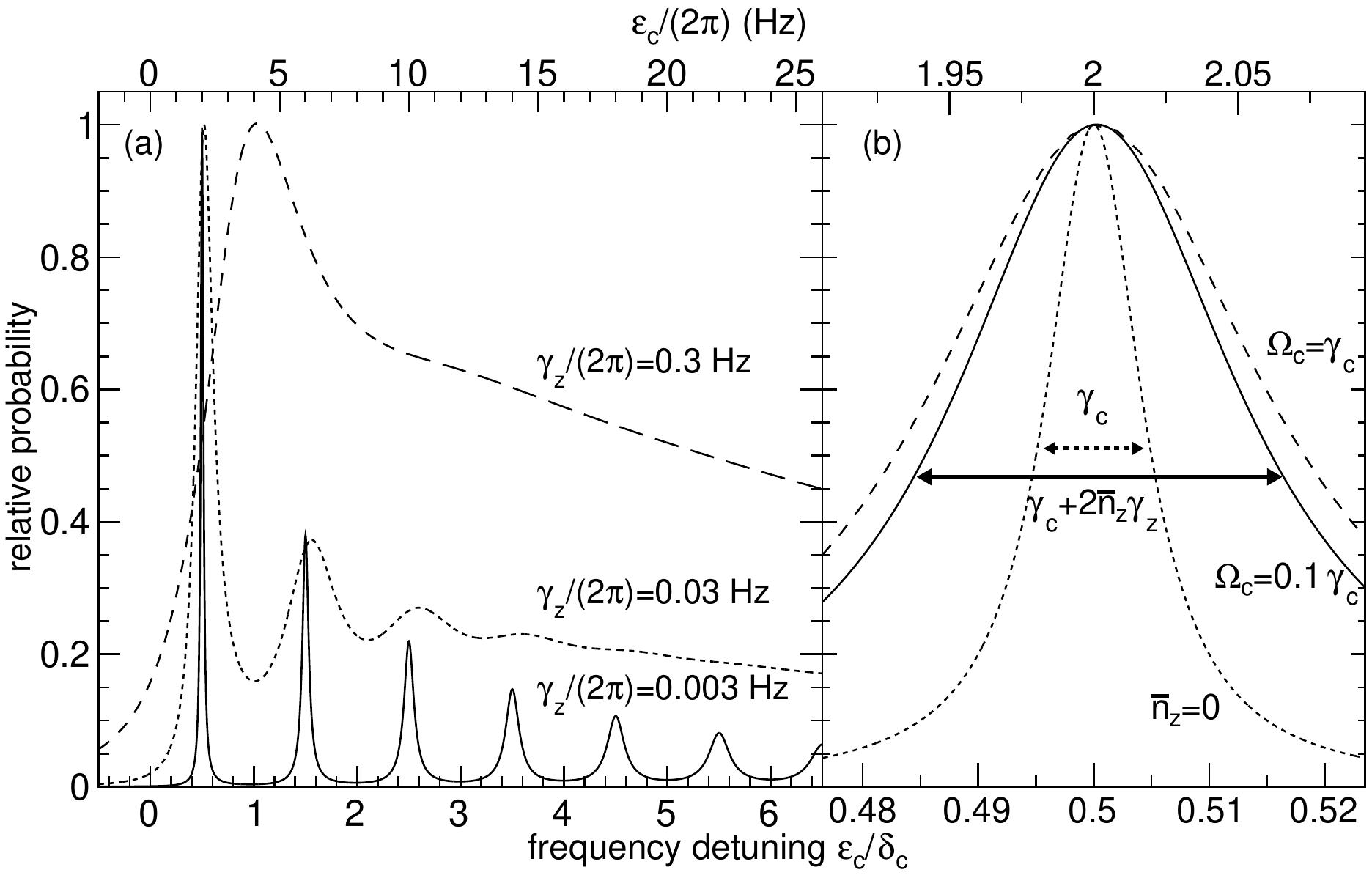}
    \caption{(a) Quantum cyclotron lineshape for $\bar{n}_z=10$ and a weak drive ($\Omega_c=0.1\gamma_c$) resolves the axial states as $\gamma_z$ is reduced.   (b) The $n_z=0$  peak for $\gamma_z/(2\pi)=0.003$ Hz and $\gamma_c/(2\pi)=0.03$ Hz for a weak (solid curve) and strong drive (dashed), and a $T=0$ K Lorentzian lineshape (dotted).} 
    \label{fig:CyclotronLineshape}
\end{figure}

Quantum calculations of the cyclotron lineshape demonstrate how the detector backaction can be reduced to only that from zero point motion. 
Figure \ref{fig:CyclotronLineshape}(a) shows steady-state lineshapes (Eq.~(\ref{eq:P})) for three values of the axial damping rate, $\gamma_z$, at a temperature $T=0.1$ K   and a coupling $\delta_c/(2\pi) = 4$ Hz.  For the dashed lineshape, $2\bar{n}_z\gamma_z/(2\pi)=6$ Hz does not satisfy Eq.~(\ref{eq:Condition1}) and the axial quantum states is not resolved.  For a ten times lower $\gamma_z/(2\pi) = 0.03$ Hz, the quantum structure of the axial motion manifests itself in the dotted lineshape. For another 10-fold reduction in $\gamma_z$, the solid lineshape shows completely resolved peaks. 

The extremely narrow left peak for $n_z=0$ is good news for measurement.  
Its width, $\gamma_c+2 \bar{n}_z \gamma_z$, is only about 3 times the cyclotron decay width $\gamma_c$, and much smaller than the total cyclotron linewidth (Fig.~\ref{fig:CyclotronLineshape}(b)).  More good news is that this $n_z=0$ peak is very symmetric about its center frequency -- a big help in precisely identifying the center frequency of the resonance.   The next peak to the right is for $n_z=1$, and so on. There are many peaks because  $\bar{n}_z = 10$ for $T=0.1$ K. 

The peak probability for a resonant weak drive, $\Omega_c=0.1\gamma_c$, is only $3.1 \times 10^{-4}$. However, increasing the cyclotron drive strength to $\Omega_c = \gamma_c$ (dashed curve in Fig.~\ref{fig:CyclotronLineshape}(b)) increases the excitation probability to $2.2\times 10^{-2}$ while power broadening the full linewidth from 3 to only 3.6 cyclotron decay widths. (The 300 differential equations for the vector master equation were integrated directly to time $10/\gamma_c$ for $\Omega_c = \gamma_c$ because the steady state solution applies only for $\Omega_c \ll \gamma_c$.) Stronger drives may be useful for tracking slow magnetic field drifts \cite{HarvardMagneticMoment2011}.  

The offset of the $n_z=0$ resonance from $\epsilon_c=0$ to $\epsilon_c = \delta_c/2$ is due to the zero point motion of the quantum axial oscillator. This could be measured in two ways. First, measuring this peak and its neighbor determines this offset, since these two peaks are spaced by twice the offset. Second, the shift of axial frequency $\omega_z$ to $\omega_z+\delta_c$ can be measured. 

In summary, a QND coupling of cyclotron motion to an axial detection motion evades all detector backaction in determining the cyclotron state.  However, it does not prevent detector backaction from broadening the observed cyclotron resonance lineshape to limit the accuracy that can be achieved in determining the cyclotron resonance frequency and the electron magnetic moment.  The first solution of quantum master equation for a quantum cyclotron and a harmonic detection oscillation demonstrates the possibility of circumventing all of the additional detector backaction except the small amount caused by the zero-point detection motion, despite a detector excitation spread over many states.     The extremely narrow and symmetric cyclotron resonance lineshapes that are predicted differ markedly from previous predictions.  The new approach promises to make it possible to make a test of the Standard Model's most precise prediction at the precision required  to check the intriguing discrepancy that now exists between prediction and measurement. 

This work was supported by the NSF, with  partial support of X.\ Fan from the Masason Foundation.  B.\ D'Urso made early contributions. B. D'Urso, S.\ E.\ Fayer, T.\ G.\ Myers, B.\ A.\ D.\ Sukra and G.\ Nahal provided useful comments.  

\bibliographystyle{prsty_gg}
\bibliography{ggrefs2018,NewRefs}

\end{document}